\documentclass[%
 reprint,
 amsmath,amssymb,
 aps,
]{revtex4-1}

\usepackage{graphicx}
\usepackage{dcolumn}
\usepackage{bm}

\begin{document}

\preprint{APS/123-QED}
\title{The Angular Momentum of Topologically Structured Darkness}

\author{Samuel N. Alperin}
 \affiliation{Department of Physics \& Astronomy, University of Denver, Denver, Colorado, USA}

\author{Mark E. Siemens}\email{msiemens@du.edu}%
 \affiliation{Department of Physics \& Astronomy, University of Denver, Denver, Colorado, USA}

\date{\today}

\begin{abstract}
We theoretically analyze and experimentally measure the extrinsic angular momentum contribution of topologically structured darkness found within fractional vortex beams, and show that this structured darkness can be explained by evanescent waves at phase discontinuities in the generating optic. We also demonstrate the first direct measurement of the intrinsic orbital angular momentum of light with both intrinsic and extrinsic angular momentum, and explain why the total orbital angular momenta of fractional vortices do not match the winding number of their generating phases.
\end{abstract}

\pacs{Valid PACS appear here}
\maketitle

Though it is well known that the orbital angular momentum (OAM) of light is a physical momentum that can be used to apply mechanical torque to small particles \citep{He:1995p3221,Padgett:2011p3140}, its classification as an intrinsic property of light remains a subject of discussion \cite{AONeil2002,Zambrini2006,Bliokh2015a}. A clearly intrinsic property such as photon spin is independent of the transverse position of the calculation axis, but the photon OAM is more complicated. Zambrini et al. showed that while the average OAM of a beam is invariant under arbitrary linear translations, its OAM spectrum is not \cite{Zambrini2006}. This led those authors to deem the OAM a \textit{quasi-intrinsic} property. It has also been stated that the angular momentum of a beam is intrinsic if and only if there exists a choice of axis such that the mode has no net transverse momentum \cite{AONeil2002,Bliokh2015a}. An extension of these conclusions is that to understand the natures of the intrinsic and extrinsic OAM, we must study a system in which both are present. In this letter we consider what is arguably the most natural of such systems: fractional vortex beams.

\begin{figure}
\begin{center}
  \includegraphics[width=\columnwidth]{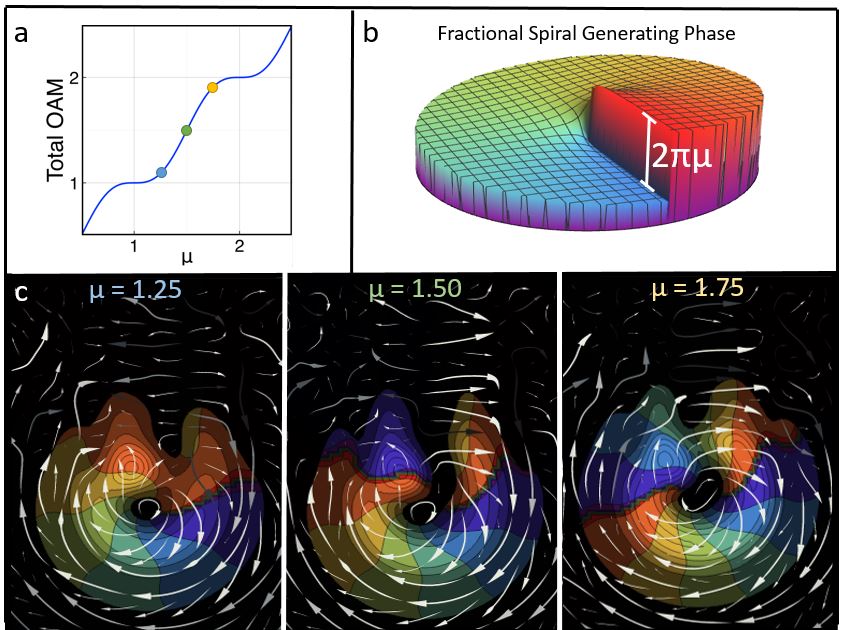}
  \caption{(a) Plot of average OAM as a function of generating spiral phase step height $\mu$, as drawn in (b). Three colored points along the curve in (a) correspond to the modes labeled in matching colors in (c). (c) Shows various theoretical fractional vortices for three different $\mu$. Intensity is represented by brightness, and phase is represented by color. The arrows show the flow of transverse Poynting vectors, which are within a scaling factor of local OAM per photon.}\label{intro}

  \end{center}
 \end{figure}

It is well known that passing a Gaussian beam through a $2\pi m$ helical phase optic imparts the beam with an OAM of $m \hbar$ for any $m\in\mathbb{Z}$. However it has only recently come to light that the OAM of \textit{fractional} vortex beams is more subtle: the mode formed by Gaussian illumination of a phase ramp of topological charge $\mu \in \mathbb{R}$ does not in general have an average OAM of $\mu \hbar$ per photon, as shown in Fig. \ref{intro}.a. Though the expectation value of the OAM of these fractional vortex modes has been calculated \citep{Gotte:2007uq} and experimentally verified \citep{Leach:2004qy,Alperin2016a}, there has been no physical explanation for the discrepancy between the topological charge of the generating phase structure and the OAM of the resulting mode. 

In this letter we show that the \textit{intrinsic} OAM of a fractional vortex beam can be directly measured, and that it matches the topological charge of the generating phase, even for non-integer $\mu$. This suggests that the fundamental action of the $2\pi\mu$ spiral phase applied to a beam is not to impart the beam with $\mu \hbar$ of total OAM, but to impart the beam with $\mu \hbar$ of \textit{intrinsic} OAM. This leads us to investigate the form and physical meaning of the other OAM component, that of \textit{extrinsic} angular momentum. We do this by separating fractional vortex modes into coherent sums of intrinsic and extrinsic components, which in the case of fractional OAM modes means separating into light and dark parts.

We start with a description of our direct optical measurement of the intrinsic OAM of light. In a prior work the authors demonstrated a technique for the quantitative measurement of total OAM by transforming the light with a cylindrical lens, recording an image with a camera at its focus, and calculating high-order moments \cite{Alperin2016a}. This OAM measurement technique exploits the fact that at the focal plane of a cylindrical lens, a camera records the intensity of the 1D Fourier transform of a mode incident on the lens. From that image the OAM can be calculated as 

\begin{equation}
\ell_{\textrm{meas}} = \frac{4 \pi}{f \lambda} \frac{\iint_{-\infty}^\infty I(x',y)_\ell x' \hspace{.02in} y \hspace{.03in} dx' \hspace{.03in} dy}{\iint_{-\infty}^\infty I(x',y)_\ell  \hspace{.03in} dx' \hspace{.03in} dy}. \label{OAMVfxy}
\end{equation}
where $I(x',y)$ is the spatially-resolved light intensity at the focal plane of the lens \cite{Alperin2016a}. However, for modes without cylindrical symmetry in the transverse plane, the cylindrical lens orientation with respect to the mode under test affects which OAM components are measured. For fractional modes, there is mirror symmetry in the transverse plane, along the line of the lateral discontinuity, and to measure the total OAM, the cylindrical lens must be oriented at $\frac{\pi}{4}$ with respect to the line of symmetry \cite{Alperin2016a}.

Any extrinsic component to the OAM can be written as a net linear transverse momentum, and thus there must be exactly one orientation of the cylindrical lens for which all of the extrinsic component is not counted in the measurement. As the intrinsic OAM is all that is left to be measured and is by definition invariant on measurement orientation, there exists exactly one orientation of the cylindrical lens for which the intrinsic OAM can be measured directly. As it is known that upon propagation the line of symmetry moves in the $\hat{\phi}$ direction, we choose the orientation of the cylindrical lens to be oriented $\frac{\pi}{2}$ radians from the orientation of the discontinuity, so as to zero the measurement to extrinsic angular momentum by mapping the transverse extrinsic momentum contribution onto the optical axis in the Fourier domain. 
 
\begin{figure}
\begin{center}
  \includegraphics[width=\columnwidth]{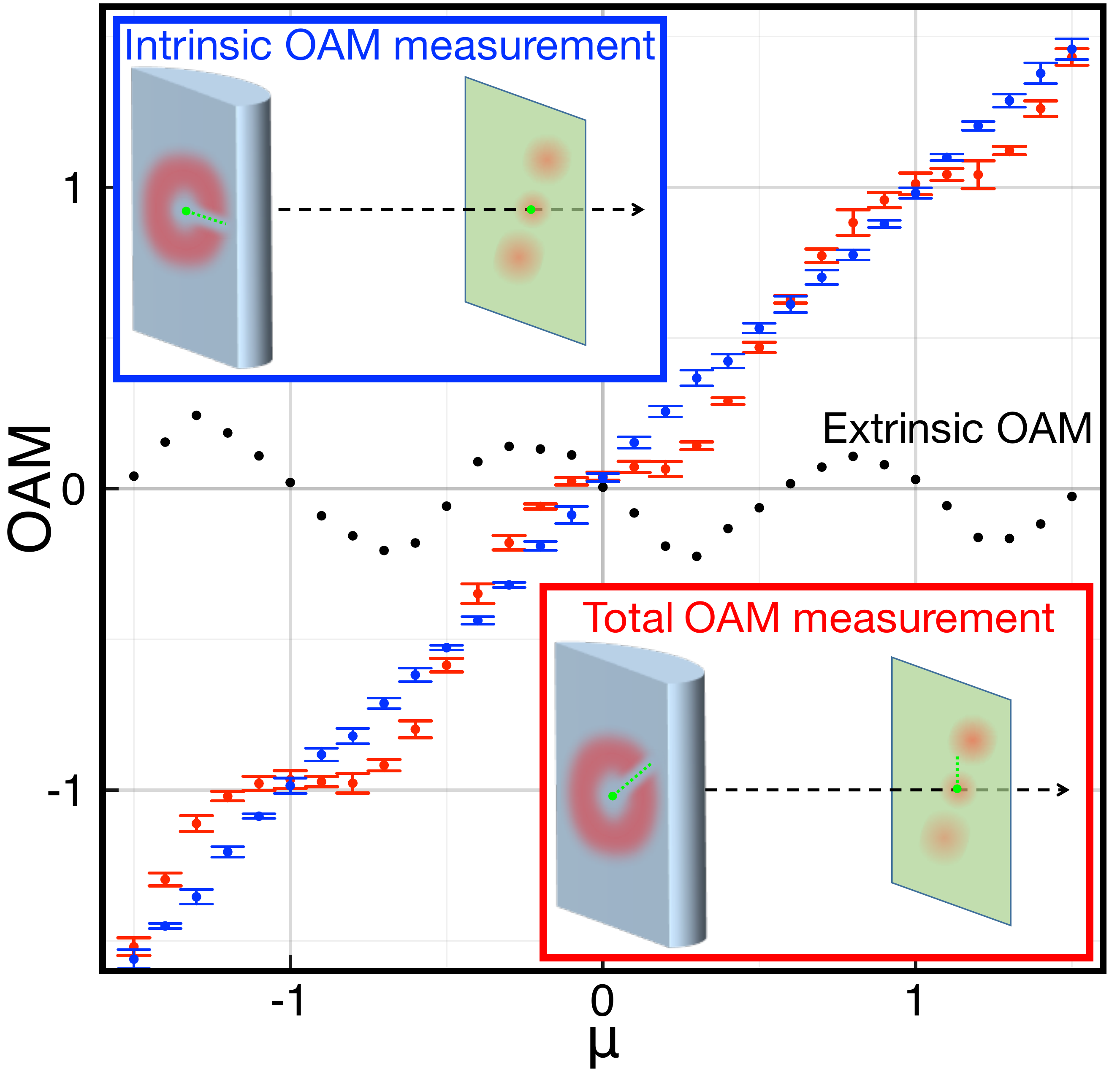}
  \caption{Blue: Direct experimental measurement of the intrinsic orbital angular momentum of fractional vortex beams generated by non-integer spiral phases of topological charge $\mu$. The blue inset shows the experimental alignment: a fractional vortex beam is input into a cylindrical lens such that the vortex chain (green dots) is perpendicular to the axis of the lens; in this case, the vortices in the chain are all focused onto the optical axis in the Fourier plane (green rectangle) and extrinsic OAM contributions are eliminated. The measured total OAM of the same fractional vortex modes is shown in red, achieved by a 45$^\circ$ rotation of the input mode, as shown in the red inset. The extrinsic OAM (black) is then calculated as the difference between the total OAM and intrinsic OAM.}\label{cyldata}
  \end{center}
  \end{figure}

The result of our intrinsic OAM measurement is shown in blue in Fig. \ref{cyldata}. This is the first direct measurement of intrinsic OAM in the presence of extrinsic OAM, and it confirms the nontrivial claim that the intrinsic OAM of fractional vortex modes goes as $\mu\hbar$. The extrinsic OAM can be calculated as the difference between the total and the intrinsic OAM, the result of which is shown in black in Fig. \ref{cyldata}.

Having shown that the intrinsic OAM of fractional vortex beams matches the topological charges of their generating phases, we now set out to understand the optical structure of these beams in connection with intrinsic and extrinsic OAM contributions.

We start by constructing the form of the fractional vortex mode. We can write the general form of a fractional vortex in terms of LG modes 

\begin{equation}
\psi_\mu=\sum_{l=-\infty}^{\infty}C_\ell \sum_{p=0}^{\infty}C_{ p}(\ell) LG_{\ell p}
\end{equation}
such that \cite{Gotte:2007uq}
\begin{equation}
C_\ell=\frac{i \left(1-e^{2 i \pi  (\mu -\lfloor \mu \rfloor )}\right) e^{i \theta  (\mu -l)-i \theta (\mu -\lfloor \mu \rfloor )}}{2 \pi  (\mu -l)}
\end{equation} 
in which $\theta$ represents the orientation angle of $\phi=0$, and thus specifies the direction of the lateral phase discontinuity in the final structure. For each $\ell$ in the sum, there is a hypergeometric response in the radial structure, as in Karimi et al \citep{Karimi:2007fj}. Restricting the mathematical expression for the hypergeometric response to physical cases, it can be simplified so that 

\begin{equation}
C_{ p}(\ell)=\frac{\ell \Gamma \left(\frac{\ell}{2}+p\right)}{2 \sqrt{p! (\ell+p)!}}.
\end{equation}
 
\begin{figure}
\begin{center}
  \includegraphics[width=\columnwidth]{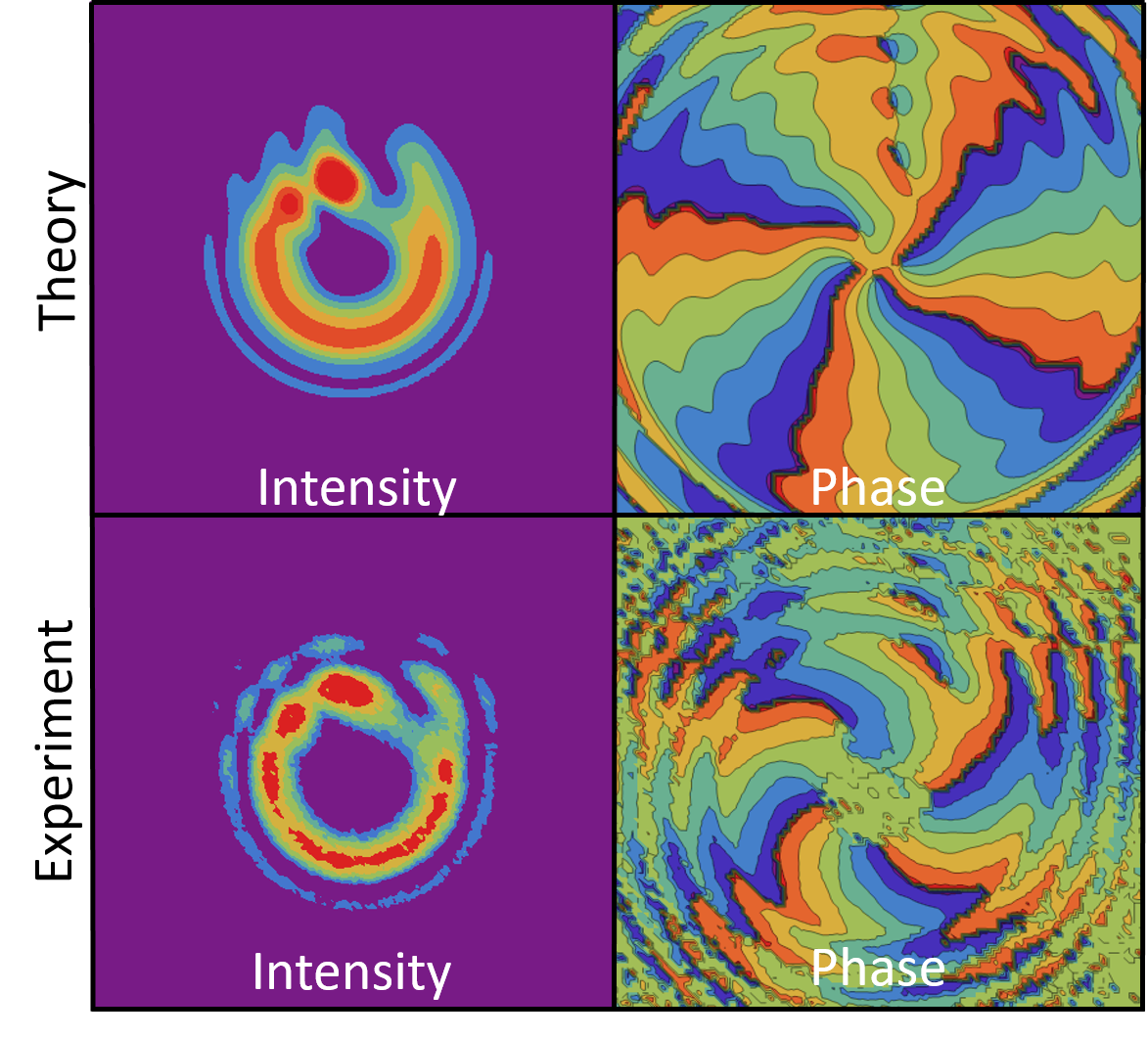}
  \caption{Intensity and phase of theoretical (top) and experimental (bottom) realizations of fractional vortex mode $\psi_{4.4}$ at a distance of $\frac{1}{20}Z_R$ from the generating optic. Notably, the vortex chain (a string of unit vortices of alternating sign) is resolved in both theoretical and experimental phase images.}\label{modeDemo}

  \end{center}

\end{figure}

 As can be seen in Fig. \ref{intro}.c, the transverse momenta of the modes have a great deal of interesting structure. Gbur elegantly showed that as $\mu$ goes continuously from one integer to the next, a radial string of unit vortices of alternating charge forms until $\mu$ is a half integer and the string extends infinitely from the center of the mode \citep{Gbur2016}. As $\mu$ continues to increase, the chain collapses towards infinity, leaving a single additional vortex which moves into the center. The vortex chain can be seen in Fig. \ref{modeDemo}, and the movement of the `additional' vortex can be seen in Fig. \ref{intro}.c: At $\mu=1.25$ there is a clearly defined, local $2\pi$ phase wrap near the mode center, while at $\mu=1.50$ the core opens up to accept the `new' vortex left by the infinite chain, and at $\mu=1.75$ there is a well defined core phase wrap of $4\pi$.
 
Fig. \ref{modeDemo} compares theoretical and experimental fields for $\psi_{4.4}$, demonstrating that our model matches experimental observations of both the intensity and phase of fractional vortex modes. In our measurements, phase was experimentally measured using phase stepping holography \cite{Yamaguchi1997,DErrico2017}. To experimentally generate fractional vortex beams, we pass a laser with a Gaussian profile through a spatial light modulator encoded with a fractional spiral phase hologram, which takes the form of a forked grating with a lateral discontinuity, and take images with a CCD a distance of $\frac{1}{20}Z_R$ beyond the modulator. The experimental results accurately resolve the dominant phase structure and the lateral stripe from the linear phase discontinuity in the generating phase, which is composed of a line of unit vortices of alternating sign that extends outward from near the center of the mode \cite{Gbur2016}.

Having built and experimentally confirmed the structure of fractional vortex beams $\psi_\mu$, we now separate them into their intrinsic and extrinsic parts, in keeping with our measurements in Fig. \ref{cyldata}. We define the intrinsic component of $\psi_\mu$ as

\begin{equation}
\widetilde{\psi_{i\mu}}(r,\phi)=|\psi_\mu(r, \tilde{\theta}+\pi)|e^{i\mu \phi}.
\end{equation}
This intrinsic component has a phase defined by the non-integer phase wrap of the generating optic, and its amplitude is a circularly symmetric wrap of the radial function of $\psi_\mu$ at the angle opposite the discontinuity, where there are no effects of the discontinuity. We are careful not to call this theoretical object a \textit{mode}, as it is not a solution to the wave equation and cannot propagate independently. However we find it conceptually useful and it is essential for the development of our understanding of its extrinsic counterpart. The intrinsic component $\widetilde{\psi_{i\mu}}$ possesses uniform average OAM of $\mu \hbar$ per photon. In order to describe the complete optical mode $\psi_\mu$, we introduce a second component, that of \textit{structured darkness}, as
\begin{equation}
\widetilde{\psi_{d\mu}}=\psi_\mu - \widetilde{\psi_{i\mu}}.
\end{equation}

\begin{figure}
\begin{center}
  \includegraphics[width=\columnwidth]{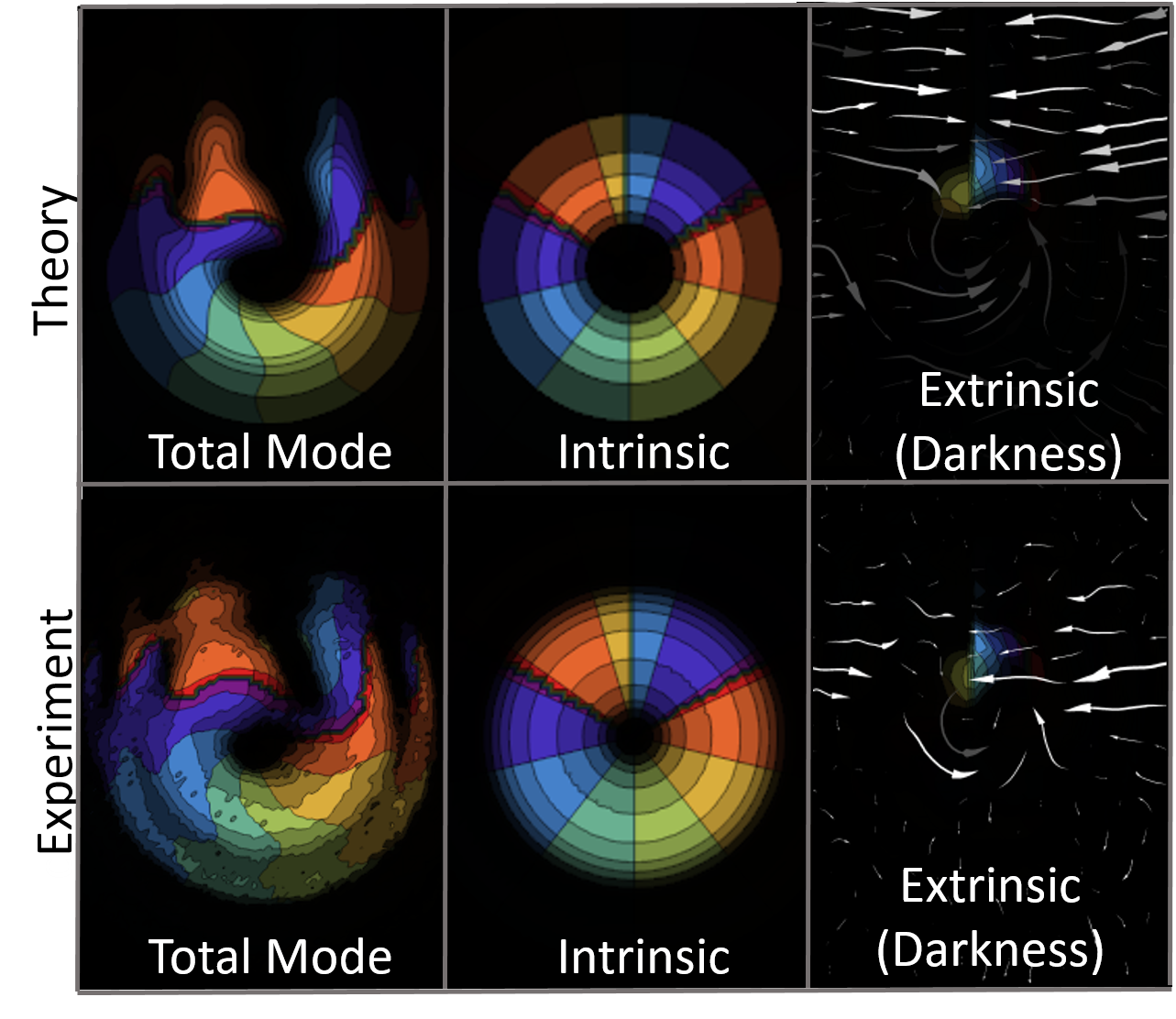}
  \caption{From left to right, theoretical (top row) and experimental (bottom row) amplitude and phase of $\psi_{1.5}$, $\widetilde{\psi_{i 1.5}}$, and $\widetilde{\psi_{d 1.5}}$. Local transverse Poynting vectors are also shown as white arrows in plots of $\widetilde{\psi_{d 1.5}}$, which demonstrate that the dominant transverse momentum is made of off axis linear momenta pointing in opposite directions, suggesting no net extrinsic contribution, which is consistent with observations in Fig. \ref{intro}.}\label{dark}

  \end{center}

\end{figure}

We can evaluate the complex form of $\widetilde{\psi_{d\mu}}$ numerically for any $\mu$. We can also perform the same operation on experimentally measured vortex beams via phase reconstruction. Fig. \ref{dark} demonstrates the match between the theory and experiment in the experimentally viable case of non-zero propagation from the generating phase optic. Both the numerical and experimental plots of the structured darkness associated with $\psi_{1.5}$ demonstrate that nearly all measurable transverse momentum takes the form of off axis, linear momentum which cancel and thus form a net zero extrinsic contribution. This is consistent with our result shown in Fig. \ref{cyldata}: the extrinsic angular momentum is zero for half-integer $\mu$. Similar analysis of modes with other values of $\mu$ were also consistent with Fig. \ref{cyldata}.

We now discuss a physical interpretation of structured darkness. It has been shown that the dark singularity at the center of an integer vortex beam, generated by the Gaussian excitation of a spiral phase, is the result of depletion of the Gaussian into non-propagating evanescent waves wherever the propagation angle of component plane waves is too high to be physical \cite{Roux2003}. Similarly, we claim that the form of the \textit{structured darkness} of a fractional vortex beam at $z=0$ represents that of evanescent waves excited along the lateral phase discontinuity of the generating phase optic.

 \begin{figure}
\begin{center}
  \includegraphics[width=\columnwidth]{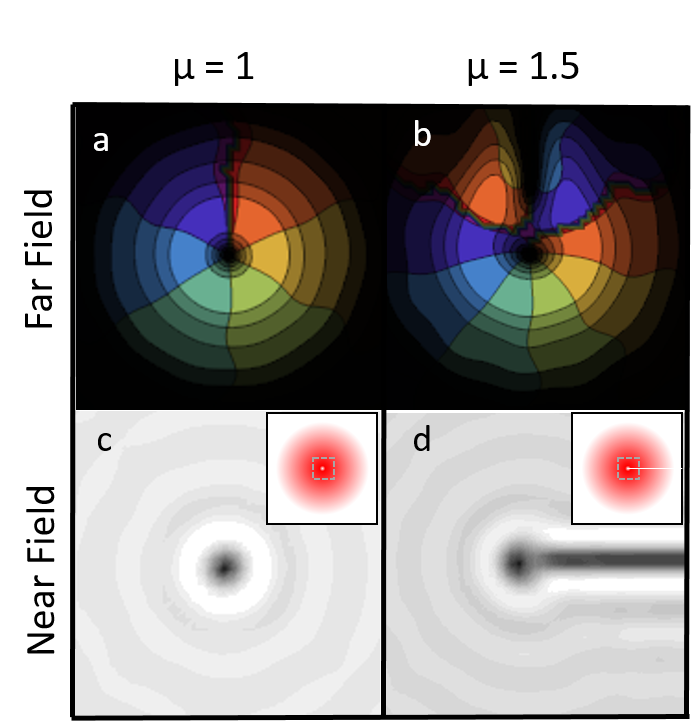}
  \caption{(Top) Far field amplitude and phase of fractional vortex beams, calculated as the propagating part of beams with Gaussian amplitude and fractional $2\pi \mu$ spiral phase. This shows that the dark stripe seen in any fractional vortex beam represents the region where the field was non-propagating. (Bottom) Near field amplitude of beam centers of integer and fractional modes generated as the Gaussian excitations of spiral phases, showing that the lateral discontinuity of fractional beams comes from depletion into evanescent waves at the surface of the generating phase optic. The orientation of the dark stripe rotates by $\frac{\pi}{2}$ over propagation to the far field. }\label{actuallylast}
  \end{center}  

  \end{figure}
To prove that the structured darkness of fractional vortex modes represents the regions depleted by the excitation of evanescent waves at the generating phase optic, we demonstrate that we can numerically generate a fractional mode by calculating the propagating, non-evanescent part of a mode with Gaussian amplitude and fractional $2\pi \mu$ spiral phase. This is done by taking the Fourier transform of the helically phased Gaussian, applying a circular aperture, and inverse Fourier transforming back. In the case that the aperture blocks only non-propagating evanescent waves, we can calculate the near field of the propagating field. By selecting a much smaller aperture in k-space, the far field of the mode can be calculated because this is mathematically equivalent to spatial filtering. Fig. \ref{actuallylast}.a-b shows the result of such far field calculations for $\mu=1$ and $\mu=1.5$. Both cases match the expectation in the far field in both phase and amplitude, without any a priori knowledge of the Laguerre-Gaussian compositions, or even of the form of any mode other than the input Gaussian. Fig. \ref{actuallylast}.c-d shows the amplitudes of the near field results for the same values of $\mu$, zoomed in to the beam center. Even in the near field, there is a clearly defined lateral singularity along the discontinuity of the generating phase. Given that this behavior is apparent with only the assumption that evanescent waves do not propagate, we conclude that the region of structured darkness seen beyond the near field represents the effect of evanescent structure at the surface of the phase optic on the propagating mode. While this is somewhat counterintuitive as it is well known that evanescent waves do not propagate, it is clear that their structure can influence that of a surrounding, propagating field.

The authors thank A.A. Voitiv for assisting in data acquisition, J.M. Knudsen, J.T. Gopinath and R.D. Niederriter for helpful discussions, and acknowledge financial support from the National Science Foundation (1509733, 1553905).


\bibliography{library} 
\bibliographystyle{ieeetr}
\end{document}